\documentstyle[aps,prd]{revtex}

\def \ep {\varepsilon}
\def \c {\mbox{curl}\,}
\def \d {\mbox{D}}
\def \ts {\textstyle}
\def \rd {\displaystyle{\cdot}}

\begin{document}
\tighten

\title{Linearisation instability of gravity waves?}

\author{Roy Maartens}

\address{School of Mathematical Studies, Portsmouth University,
Portsmouth PO1 2EG, England\\
and Department of Mathematics and Applied Mathematics, University of
Natal, Durban 4001, South Africa}


\maketitle

\begin{abstract}

Gravity waves in irrotational dust spacetimes are 
characterised by nonzero magnetic Weyl tensor $H_{ab}$. In
the linearised theory, the divergence of $H_{ab}$ is set to
zero. 
Recently Lesame et al. [Phys. Rev. D {\bf 53}, 738
(1996)] presented an argument to show 
that, in the exact nonlinear theory,
$\mbox{div}\,H=0$ forces $H_{ab}=0$, thus implying a 
linearisation instability for gravity waves interacting with
matter. 
However a sign error in the equations invalidates their
conclusion.
Bianchi type V spacetimes are shown to include examples 
with $\mbox{div}\,H=0\neq H_{ab}$.
An improved covariant
formalism is used to show that in 
a generic irrotational dust spacetime, the covariant constraint
equations are preserved under evolution.
It is shown elsewhere that $\mbox{div}\,H=0$ does not generate further
conditions.


\end{abstract}

\section{Introduction}

Irrotational dust spacetimes have been widely studied, in
particular as models for the late universe, and as arenas for
the evolution of density perturbations and gravity wave
perturbations. In linearised theory, i.e. where the irrotational
dust spacetime is close to a Friedmann--Robertson--Walker dust
spacetime, gravity wave perturbations are usually
characterised by 
transverse traceless tensor modes. 
In terms of the covariant and gauge--invariant 
perturbation formalism initiated by Hawking \cite{h}
and developed by Ellis and Bruni \cite{eb},
these perturbations are described by the
electric and magnetic Weyl tensors, given respectively by
\begin{equation}
E_{ab}=C_{acbd}u^c u^d\,,\quad H_{ab}={\ts{1\over2}}\eta_{acde}u^e
C^{cd}{}{}_{bf}u^f
\label{eh}
\end{equation}
where $C_{abcd}$ is the Weyl tensor, $\eta_{abcd}$ is the 
spacetime permutation tensor, and $u^a$ is the dust four--velocity.
In the so--called `silent universe' case 
$H_{ab}=0$, no information is exchanged between neighbouring
particles, also in the exact nonlinear case. Gravity wave
perturbations require nonzero $H_{ab}$, which is 
divergence--free in the linearised case \cite{led}, \cite{he},
\cite{b}.

A crucial question for the analysis of gravity waves 
interacting with matter is whether
the properties of the linearised perturbations are 
in line with those of the exact nonlinear theory. 
Lesame et al. \cite{led} used the covariant formalism 
and then specialised to a shear tetrad, in order to
study this question. They concluded that in the nonlinear case,
the only solutions with $\mbox{div}\,H=0$ are those with $H_{ab}=0$
--- thus indicating a linearisation instability, with potentially
serious implications for standard analyses of gravity waves, as 
pointed out in \cite{m}, \cite{ma}.
It is shown here that the argument of \cite{led} 
does not in fact prove that
$\mbox{div}\,H=0$ implies
$H_{ab}=0$.
The error in \cite{led} is traced to an incorrect sign
in the Weyl tensor decomposition (see below).\footnote{The authors of
\cite{led} are in agreement about the error and its implication 
(private communication).}

The same covariant formalism is used here, but with modifications 
that
lead to simplification and greater clarity. This improved
covariant formalism renders the equations more transparent, and
together with the new identities derived via the formalism,
it facilitates a fully covariant analysis, 
not requiring 
lengthy tetrad calculations such as those
used in \cite{led}. 
The improved formalism
is presented in Section II, and the identities that are crucial for
covariant analysis are given in the appendix.

In Section III, a covariant derivation is given to show 
that {\em in the generic case of irrotational dust
spacetimes, the constraint equations are
preserved under evolution.}
A by--product of the 
argument is the identification of the
error in \cite{led}. 
In a companion paper \cite{mel},
we use the covariant formalism of Section III 
to show that when $\mbox{div}\,H=0$,
no further conditions are generated. In particular, $H_{ab}$ {\em is 
not forced to vanish, and there is not a linearisation instability.} 
A specific example is presented in
Section IV, where
it is shown that Bianchi type V spacetimes 
include cases in which
$\mbox{div}\,H=0$ but $H_{ab}\neq0$.

\section{The covariant formalism for propagation and 
constraint equations}

The notation and conventions are based on 
those of \cite{led}, \cite{e1};
in particular $8\pi G=1=c$, round brackets enclosing indices
denote symmetrisation and square brackets denote
anti--symmetrisation. Curvature tensor conventions are given in
the appendix.

Considerable simplification and streamlining results from
the following definitions: the projected permutation tensor 
(compare \cite{e3}, \cite{mes}),
\begin{equation}
\ep_{abc}=\eta_{abcd}u^d
\label{d1}
\end{equation}
the projected, symmetric and trace--free part of a tensor,
\begin{equation}
S_{<ab>}=h_a{}^c h_b{}^d S_{(cd)}-
{\ts{1\over3}}S_{cd}h^{cd} h_{ab}
\label{d2}
\end{equation}
where $h_{ab}=g_{ab}+u_au_b$ is the spatial projector
and $g_{ab}$ is the metric,
the projected spatial covariant derivative (compare \cite{e2}, 
\cite{eb}, \cite{mes}),
\begin{equation}
\d_a S^{c\cdots d}{}{}{}{}_{e\cdots f}=h_a{}^b h^c{}_p \cdots
h^d{}_q h_e{}^r \cdots h_f{}^s \nabla_b
S^{p\cdots q}{}{}{}_{r\cdots s}
\label{d3}
\end{equation}
and the covariant spatial curl of a tensor,
\begin{equation}
\c S_{ab}=\ep_{cd(a}\d^c S_{b)}{}^d
\label{d4}
\end{equation}
Note that  
$$
S_{ab}=S_{(ab)}\quad\Rightarrow\quad\c S_{ab}=\c S_{<ab>}
$$
since $\c(fh_{ab})=0$ for any $f$. 
The covariant spatial divergence of $S_{ab}$ is 
$$(\mbox{div}\,S)_a=\d^b S_{ab}$$
The covariant spatial curl of a vector is
$$
\c S_a=\ep_{abc}\d^bS^c
$$

Covariant analysis of propagation and constraint equations
involves frequent use of a number of algebraic and differential
identities governing the above quantities. In particular, one
requires commutation rules for spatial and time derivatives.
The necessary identities are collected for convenience in the 
appendix, which includes a simplification of 
known results and a number of new results.

The Einstein, Ricci and Bianchi equations may be covariantly split
into propagation and constraint equations \cite{e1}.
The propagation equations given in
\cite{led} for irrotational dust are simplified by the present
notation, and become
\begin{eqnarray}
\dot{\rho}+\Theta\rho &=& 0 
\label{p1}\\
\dot{\Theta}+{\ts{1\over3}}\Theta^2 &=& -{\ts{1\over2}}\rho
-\sigma_{ab}\sigma^{ab}
\label{p2}\\
\dot{\sigma}_{ab}+{\ts{2\over3}}\Theta\sigma_{ab}+\sigma_{c<a}
\sigma_{b>}{}^c &=& -E_{ab} 
\label{p3}\\
\dot{E}_{ab}+\Theta E_{ab}-3\sigma_{c<a}E_{b>}{}^c &=&
\c H_{ab}-{\ts{1\over2}}\rho\sigma_{ab} 
\label{p4}\\
\dot{H}_{ab}+\Theta H_{ab}-3\sigma_{c<a}H_{b>}{}^c &=& -\c E_{ab}
\label{p5}
\end{eqnarray}
while the constraint equations become
\begin{eqnarray}
\d^b\sigma_{ab} &=& {\ts{2\over3}}\d_a \Theta
\label{c1}\\
\c \sigma_{ab}&=& H_{ab}
\label{c2}\\
\d^b E_{ab} &=& {\ts{1\over3}}\d_a \rho +
\ep_{abc}\sigma^b{}_d H^{cd}
\label{c3}\\
\d^b H_{ab} &=& -\ep_{abc}\sigma^b{}_d E^{cd}
\label{c4}
\end{eqnarray}
A dot denotes a covariant derivative along $u^a$, $\rho$ is the
dust energy density, $\Theta$ its rate of
expansion, and $\sigma_{ab}$ its shear. Equations (\ref{p4}),
(\ref{p5}), (\ref{c3}) and (\ref{c4}) display the analogy with
Maxwell's theory. The FRW case is covariantly characterised by
$$
\d_a\rho=0=\d_a\Theta\,,\quad\sigma_{ab}=E_{ab}=H_{ab}=0
$$
and in the linearised case of an almost FRW spacetime, these gradients
and tensors are first order of smallness.

The dynamical fields in these equations are the scalars $\rho$ and
$\Theta$, and the 
tensors $\sigma_{ab}$,
$E_{ab}$ and $H_{ab}$, which all satisfy $S_{ab}=S_{<ab>}$. The
metric $h_{ab}$ of the spatial
surfaces orthogonal to $u^a$ is implicitly
also involved in the equations as a dynamical field. Its propagation
equation is simply the identity $\dot{h}_{ab}=0$, 
and its constraint equation is the identity $\d_a h_{bc}=0$ --
see (\ref{a4}). The Gauss--Codacci equations for the Ricci curvature
of the spatial surfaces \cite{e1}
\begin{eqnarray}
R^*_{ab}-{\ts{1\over3}}R^*h_{ab} &=& -\dot{\sigma}_{ab}-\Theta
\sigma_{ab} \nonumber\\
R^* &=&-{\ts{2\over3}}\Theta^2+\sigma_{ab}\sigma^{ab}+2\rho \label{r1}
\end{eqnarray}
have not been included, since the curvature is algebraically 
determined by the other fields,
as follows from (\ref{p3}):
\begin{equation}
R^*_{ab}=E_{ab}-{\ts{1\over3}}\Theta\sigma_{ab}+\sigma_{ca}
\sigma_b{}^c+{\ts{2\over3}}\left(\rho-{\ts{1\over3}}\Theta^2\right)
h_{ab}
\label{r2}\end{equation}
The contracted Bianchi identities for the 3--surfaces \cite{e1}
$$
\d^b R^*_{ab}={\ts{1\over2}}\d_a R^*
$$
reduce to the Bianchi constraint (\ref{c3}) on using (\ref{c1}),
(\ref{c2}) and the identity (\ref{a13}) in (\ref{r1}) and
(\ref{r2}). Consequently, these identities do not impose any new
constraints.

By the constraint (\ref{c2}), one can in principle eliminate $H_{ab}$.
However, this leads to second--order derivatives in the propagation
equations (\ref{p4}) and (\ref{p5}). It seems preferable to maintain
$H_{ab}$ as a basic field. 

One interesting use of (\ref{c2}) is in
decoupling the shear from the Weyl tensor. 
Taking the time derivative of
the shear propagation equation (\ref{p3}), using the propagation
equation (\ref{p4}) and the constraint (\ref{c2}), together with
the identity (\ref{a16}), one gets
\begin{eqnarray}
&&-\d^2\sigma_{ab}+\ddot{\sigma}_{ab}+{\ts{5\over3}}\Theta
\dot{\sigma}_{ab}-{\ts{1\over3}}\dot{\Theta}\sigma_{ab}+
{\ts{3\over2}}\d_{<a}\d^c\sigma_{b>c} \nonumber\\
&&{}=4\Theta\sigma_{c<a}\sigma_{b>}{}^c+6\sigma^{cd}\sigma_{c<a}
\sigma_{b>d}-2\sigma^{de}\sigma_{de}h_{c<a}\sigma_{b>}{}^c+
4\sigma_{c<a}\dot{\sigma}_{b>}{}^c
\label{s}\end{eqnarray}
where $\d^2=\d^a \d_a$ is the covariant Laplacian. 
This is {\em the exact nonlinear generalisation of the linearised
wave equation for shear perturbations} derived in \cite{he}.
In the linearised 
case, the right hand side of (\ref{s}) vanishes, leading to a
wave equation governing the propagation of shear perturbations in
an almost FRW dust spacetime:
$$
-\d^2\sigma_{ab}+\ddot{\sigma}_{ab}+{\ts{5\over3}}\Theta
\dot{\sigma}_{ab}-{\ts{1\over3}}\dot{\Theta}\sigma_{ab}+
{\ts{3\over2}}\d_{<a}\d^c\sigma_{b>c} \approx 0
$$

As suggested by comparison of (\ref{c2}) and (\ref{c4}), and
confirmed by the identity (\ref{a14}), div~curl is {\em not} zero,
unlike its Euclidean vector counterpart. Indeed, the divergence of
(\ref{c2}) reproduces (\ref{c4}), on using the (vector) curl
of (\ref{c1}) and
the identities 
(\ref{a2}), (\ref{a8}) and (\ref{a14}):
\begin{equation}
\mbox{div (\ref{c2}) and curl (\ref{c1})}\quad\rightarrow\quad
\mbox{(\ref{c4})}
\label{i1}\end{equation}
Further
differential relations amongst the propagation and constraint
equations are
\begin{eqnarray}
\mbox{curl (\ref{p3}) and (\ref{c1}) and (\ref{c2}) and 
(\ref{c2})$^{\rd}$}\quad
& \rightarrow & \quad\mbox{(\ref{p5})} \label{i2}\\
\mbox{grad (\ref{p2}) and div (\ref{p3}) and (\ref{c1}) and
(\ref{c1})$^{\rd}$ and
(\ref{c2})}\quad & \rightarrow & \quad \mbox{(\ref{c3})} \label{i3}
\end{eqnarray}
where the identities (\ref{a7}), (\ref{a11.}), (\ref{a13}),
(\ref{a13.}) and (\ref{a15}) have been used.

Consistency
conditions may arise 
to preserve the constraint equations under
propagation along $u^a$ \cite{led}, \cite{he}.
In the general
case, i.e. without imposing any assumptions about 
$H_{ab}$ or other quantities, the constraints are 
preserved under evolution. 
This is shown in the next section, and forms the
basis for analysing special cases, such as
$\mbox{div}\,H=0$.

\section{Evolving the constraints: general case}

Denote the constraint equations (\ref{c1}) --- (\ref{c4}) by
${\cal C}^A=0$, where
$$
{\cal C}^A=\left(\d^b\sigma_{ab}-{\ts{2\over3}}\d_a\Theta\,,\,
\c\sigma_{ab}-H_{ab}\,,\,\cdots\right)
$$
and $A={\bf 1},\cdots, {\bf 4}$.
The evolution of ${\cal C}^A$ along $u^a$ leads to a
system of equations $\dot{{\cal C}}^A={\cal F}^A
({\cal C}^B)$, where ${\cal F}^A$ do not contain
time derivatives, since these are eliminated via the propagation
equations and suitable identities. Explicitly, one obtains after
lengthy calculations the following:
\begin{eqnarray}
\dot{{\cal C}}^{\bf 1}{}_a&=&-\Theta{\cal C}^{\bf 1}{}_a+2\ep_a{}^{bc}
\sigma_b{}^d{\cal C}^{\bf 2}{}_{cd}-{\cal C}^{\bf 3}{}_a
\label{pc1}\\
\dot{{\cal C}}^{\bf 2}{}_{ab}&=&-\Theta{\cal C}^{\bf 2}{}_{ab}
-\ep^{cd}{}{}_{(a}\sigma_{b)c}{\cal C}^{\bf 1}{}_d
\label{pc2}\\
\dot{{\cal C}}^{\bf 3}{}_a&=&-{\ts{4\over3}}\Theta{\cal C}^{\bf 3}{}_a
+{\ts{1\over2}}\sigma_a{}^b{\cal C}^{\bf 3}{}_b-{\ts{1\over2}}\rho
{\cal C}^{\bf 1}{}_a  \nonumber\\
&&{}+{\ts{3\over2}}E_a{}^b{\cal C}^{\bf 1}{}_b 
-\ep_a{}^{bc}E_b{}^d{\cal C}^{\bf 2}
{}_{cd}+{\ts{1\over2}}\c{\cal C}^{\bf 4}{}_a
\label{pc3}\\
\dot{{\cal C}}^{\bf 4}{}_a&=&-{\ts{4\over3}}\Theta{\cal C}^{\bf 4}{}_a
+{\ts{1\over2}}\sigma_a{}^b{\cal C}^{\bf 4}{}_b
 \nonumber\\
&&{}+{\ts{3\over2}}H_a{}^b{\cal C}^{\bf 1}{}_b 
-\ep_a{}^{bc}H_b{}^d{\cal C}^{\bf 2}
{}_{cd}-{\ts{1\over2}}\c{\cal C}^{\bf 3}{}_a
\label{pc4}
\end{eqnarray}
For completeness, the following list of equations used in the
derivation is given:\\
Equation
(\ref{pc1}) requires (\ref{a7}), (\ref{a11.}), (\ref{p2}), (\ref{p3}),
(\ref{c1}), (\ref{c2}), (\ref{c3}), (\ref{a13}) -- where (\ref{a13})
is needed to eliminate the following term from the right hand side
of (\ref{pc1}):
\begin{eqnarray*}
&&\ep_{abc}\sigma^b{}_d\,\c\sigma^{cd}
-\sigma^{bc}\d_a \sigma_{bc}\\
&&{}+\sigma^{bc}
\d_c \sigma_{ab}+{\ts{1\over2}}\sigma_{ac}\d_b\sigma^{bc} \equiv0
\end{eqnarray*}
Equation
(\ref{pc2}) requires (\ref{a15}), (\ref{p3}), (\ref{p5}), (\ref{c1}),
(\ref{c2}), (\ref{a3.}) -- where (\ref{a3.}) is needed to eliminate
the following term from the right hand side of (\ref{pc2}):
$$
\ep_{cd(a}\left\{\d^c\left[\sigma_{b)}{}^e\sigma^d{}_e\right]+
\d^e\left[\sigma_{b)}{}^d\sigma^c{}_e\right]\right\}\equiv0
$$
Equation
(\ref{pc3}) requires (\ref{a11.}), (\ref{p1}), (\ref{p4}), (\ref{p5}),
(\ref{a14}), (\ref{a3}), (\ref{c1}), (\ref{c3}), (\ref{c4}),
(\ref{a13}) -- where (\ref{a13}) is needed to eliminate the
following term from the right hand side of (\ref{pc3}):
\begin{eqnarray*}
&& {\ts{1\over2}}\sigma_{ab}\d_c E^{bc}
+\ep_{abc}E^b{}_d\, \c\sigma^{cd}\\
& &{}+\ep_{abc}\sigma^b{}_d
\,\c E^{cd}
+{\ts{1\over2}}E_{ab}\d_c\sigma^{bc}+E^{bc}\d_b\sigma_{ac}\\
& &{}+\sigma^{bc}\d_b E_{ac}-
\d_a\left(\sigma^{bc}E_{bc}\right)\equiv 0
\end{eqnarray*}
Equation
(\ref{pc4}) requires (\ref{a11.}), (\ref{p3}), (\ref{p4}), (\ref{p5}),
(\ref{a14}), (\ref{a13}), (\ref{c1}), (\ref{c2}), (\ref{c3}),
(\ref{c4}).

In \cite{led}, a sign error in the Weyl tensor decomposition
(\ref{a5}) led to spurious consistency conditions arising from
the evolution of (\ref{c1}), (\ref{c2}). The evolution
of the Bianchi constraints (\ref{c3}), (\ref{c4})
was not considered in \cite{led}.

Now suppose that the constraints
are satisfied on an initial spatial surface $\{t=t_0\}$, i.e.
\begin{equation}
{\cal C}^A\Big|_{t_0}=0
\label{i}\end{equation}
where
$t$ is proper time along the dust worldlines. Then by 
(\ref{pc1}) -- (\ref{pc4}), it follows that the 
constraints are satisfied for all time, since ${\cal C}^A=0$ is 
a solution for the given initial data. Since the system is linear,
this solution is unique.

This establishes that the constraint equations are preserved under 
evolution. However, it does not prove existence of solutions to
the constraints in the generic case
--- only that if solutions exist, then they evolve
consistently. The question of existence is currently under 
investigation. One would like to show explicitly how a metric
is constructed from given initial data in the covariant formalism.
This involves in particular considering whether the 
constraints generate new constraints, i.e. whether they are
integrable as they stand, or whether there are implicit
integrability conditions. The relation (\ref{i1}) is part of the
answer to this question, in that it shows how, within any
$\{t=\mbox{ const}\}$ surface, the constraint ${\cal C}^{\bf 4}$
is satisfied if ${\cal C}^{\bf 1}$ and ${\cal C}^{\bf 2}$ are
satisfied. Specifically, (\ref{i1}) shows that
\begin{equation}
{\cal C}^{\bf 4}{}_a={\ts{1\over2}}\c{\cal C}^{\bf 1}
{}_a-\d^b{\cal C}^{\bf 2}{}_{ab}
\label{i4}\end{equation}
Hence, if one takes ${\cal C}^{\bf 1}$ as determining 
$\mbox{grad}\,\Theta$,
${\cal C}^{\bf 2}$ as defining $H$ and ${\cal C}^{\bf 3}$ as
determining $\mbox{grad}\,\rho$, the constraint equations are 
consistent with each other because ${\cal C}^{\bf 4}$ then follows.
Thus if there exists a solution to the constraints on
$\{t=t_0\}$, then it is consistent and it evolves consistently.

In the next section, Bianchi type V spacetimes are shown to provide
a concrete example of existence and consistency in the case
$$
\mbox{div}\,E\neq 0\neq\c E\,,\quad\mbox{div}\,H=0\neq\c H\,,\quad
\mbox{grad}\,\rho=0=\mbox{grad}\,\Theta
$$

\section{Spacetimes with $\mbox{div}\,H=0\neq H$}

Suppose now that the magnetic Weyl tensor is divergence--free, a
necessary condition for gravity waves: 
\begin{equation}
\mbox{div}\,H=0\quad\Leftrightarrow\quad [\sigma,E]=0
\label{dh}\end{equation}
where $[S,V]$ is the index--free notation for the covariant commutator
of tensors [see (\ref{a2})], and the equivalence follows from
the constraint (\ref{c4}). 
Using the covariant 
formalism of Section III, it can be shown \cite{mel} that (\ref{dh})
is preserved under evolution without generating further conditions.
In particular, (\ref{dh}) does not force $H_{ab}=0$ -- as shown by
the following explicit example.

First note that by (\ref{r2}) and (\ref{dh}):
$$
R^*_{ab}={\ts{1\over3}}R^*h_{ab}\quad\Rightarrow\quad
[\sigma,R^*]=0\quad\Rightarrow\quad\mbox{div}\,H=0
$$
i.e., {\em irrotational dust spacetimes
have $\mbox{div}\,H=0$ if $R^*_{ab}$ is isotropic.}

Now the example arises from the class of irrotational spatially 
homogeneous spacetimes,
comprehensively analysed and classified by Ellis and MacCallum
\cite{em}. 
According to Theorem 7.1 of \cite{em}, the only non--FRW 
spatially homogeneous spacetimes
with $R^*_{ab}$ isotropic are Bianchi type I and 
(non--axisymmetric) Bianchi type V. The former have $H_{ab}=0$.
For the latter, using
the shear eigenframe $\{{\bf e}_a\}$ of \cite{em}
\begin{equation} 
\sigma_{ab} = \sigma_{22}\,\mbox{diag}(0,0,1,-1) \label{b0}
\end{equation}
Using (\ref{r1}) and (\ref{r2}) with (\ref{b0}), one
obtains
\begin{eqnarray}
E_{ab} &=& {\textstyle{1\over3}}\Theta\sigma_{ab}-\sigma_{c<a}
\sigma_{b>}{}^c \nonumber\\
&=&{\textstyle{1\over3}}
\sigma_{22}\,\mbox{diag}\left(0,2\sigma_{22},\Theta-\sigma_{22},
-\Theta-\sigma_{22}\right) \label{b0'}
\end{eqnarray}
in agreement with \cite{em}.\footnote{Note that
$E_{ab}$ in \cite{em} is the negative of $E_{ab}$ defined 
in (\ref{eh}).}

The tetrad forms of div and curl 
for type V are (compare \cite{vu}):
\begin{eqnarray}
\d^b S_{ab}&=&\partial_b S_a{}^b-
3a^b S_{ab} \label{b2}\\
\c S_{ab} &=& \ep_{cd(a}\partial^c 
S_{b)}{}^d+\ep_{cd(a}S_{b)}{}^c a^d \label{b3}
\end{eqnarray}
where $S_{ab}=S_{<ab>}$, $a_b=a\delta_b{}^1$ 
($a$ is the type V Lie algebra parameter) and
$\partial_a f$ is the directional derivative of $f$
along ${\bf e}_a$. Using (\ref{b3}) and (\ref{c2}):
\begin{eqnarray}
H_{ab} &=& \c\sigma_{ab}\nonumber\\
&=&-2a\sigma_{22}\delta_{(a}{}^2\delta_{b)}{}^3
\label{b1}\end{eqnarray}
Hence:\\ {\em Irrotational Bianchi V dust spacetimes in general
satisfy} $\mbox{div}\,H=0\neq H$. 

Using (\ref{b0})---(\ref{b1}), one obtains
\begin{eqnarray}
\d^bH_{ab}&=&0  \label{v1}\\
\c H_{ab}&=& -a^2\sigma_{ab} \label{v2}\\
\c\c H_{ab}&=& -a^2H_{ab} \label{v3}\\
\d^bE_{ab} &=& -\sigma_{bc}\sigma^{bc}a_a \label{v4}\\
\c E_{ab} &=&{\textstyle{1\over3}}\Theta H_{ab} \label{v5}
\end{eqnarray}

Although (\ref{v1}) is a necessary condition for gravity waves,
it is not sufficient, and (\ref{b0'}) and (\ref{b1}) show that
$E_{ab}$ and $H_{ab}$ decay with the shear, so that
the type V solutions cannot be interpreted as gravity waves. 
Nevertheless, these solutions do establish the existence of
spacetimes with $\mbox{div}\,H=0\neq H$.       

This supplements the known result that the only spatially homogeneous 
irrotational dust spacetimes with $H_{ab}=0$ are FRW, Bianchi types
I and VI$_{-1}$ $(n^a{}_a=0)$, and Kantowski--Sachs \cite{bmp}.
When $H_{ab}=0$, (\ref{b0}) and (\ref{b1}) show that $\sigma_{ab}=0$, 
in which case the type V solution reduces to FRW.\\

A final remark concerns the special case $H_{ab}=0$, i.e. the
silent universes. The considerations of this paper show that the
consistency analysis of silent universes undertaken in \cite{lde}
needs to be re--examined. This is a further topic currently under
investigation. It seems likely that the silent universes, in the
full nonlinear theory, are {\em not} in general consistent.

\acknowledgements 
Thanks to the referee for very helpful comments, and
to George Ellis, William Lesame and Henk van Elst 
for very useful discussions.
This research was supported by grants from Portsmouth, Natal and
Cape Town Universities. Natal University, and especially Sunil
Maharaj, provided warm hospitality while part of this research
was done.

\appendix
\section*{Covariant Identities}

In this appendix 
$$
S_{ab}=S_{<ab>}\quad\mbox{and}\quad V_{ab}=V_{<ab>}
$$

Using the properties of $\eta_{abcd}$ (see \cite{e1}) and $h_{ab}$,
one can derive
\begin{eqnarray}
\ep^{abc}\ep_{def} &=& 3!\,h^{[a}{}_d h^b{}_e h^{c]}{}_{f}
\label{a1}\\
\ep_{abc}S^b{}_dV^{cd}&=&{\ts{1\over2}}\ep_{abc}
\left[ S,V \right]^{bc}
\label{a2}
\end{eqnarray}
where $[S,V]$ is the covariant commutator of rank--2 tensors. 
Using a tetrad that diagonalises $S_{ab}$, one may prove the
further covariant identities:
\begin{eqnarray}
\ep_{abc}S^b{}_pS^p{}_q V^{cq} &=& -S_{ab}\ep^{bcd}S_c{}^pV_{dp} 
\label{a3}\\
\c(S^2)_{ab}&=&\ep_{cd(a}\d^e
\left\{S_{b)}{}^c S^d{}_e\right\}
\label{a3.}
\end{eqnarray}
where $(S^2)_{ab}=S_a{}^cS_{cb}$.
Furthermore
\begin{equation}
\d_a h_{bc}=0=\d_a \ep_{bcd}\,,\quad\quad 
\dot{h}_{ab}=0=\dot{\ep}_{abc}
\label{a4}
\end{equation}
where the time derivative identities depend on $\dot{u}_a=0$, while
the spatial derivative identities do not.

The curvature tensor is
$$
R_{abcd}=C_{abcd}+g_{a[c}R_{d]b}-g_{b[c}R_{d]a}-{\ts{1\over6}}
Rg_{abcd}
$$
where $g_{abcd}\equiv g_{ac}g_{bd}-g_{ad}g_{bc}$, and
\begin{eqnarray}
C_{abcd} &=& (g_{abpq}g_{cdrs}-\eta_{abpq}\eta_{cdrs})u^pu^rE^{qs}
\nonumber\\
& &{}-(\eta_{abpq} g_{cdrs}+g_{abpq}\eta_{cdrs})u^pu^rH^{qs}
\label{a5}
\end{eqnarray}
(see \cite{e1}, corrected here and in agreement with
\cite{b}). By the field equations
for dust, the Ricci tensor is
$$
R_{ab}\equiv R^c{}_{acb}={\ts{1\over2}}\rho(u_au_b+h_{ab})
$$
and, together with (\ref{a5}), this gives
\begin{eqnarray}
R_{abcd} &=& 2\left(h_{a[c}+u_au_{[c}\right)E_{d]b}+
2E_{a[c}\left(h_{d]b}+u_{d]}u_b\right) \nonumber\\
& &{}+2\ep_{abe}u_{[c}H_{d]}{}^e+2\ep_{cde}u_{[a}H_{b]}{}^e
\nonumber\\
 & &{} +{\ts{1\over3}}\rho\left(h_{a[c}u_{d]}u_a-h_{b[c}u_{d]}u_a
+2h_{a[c}h_{d]b}\right)
\label{a6}
\end{eqnarray}
The Ricci identities $f_{;[ab]}=0$ and
$$
2Y_{a;[bc]}=R^d{}_{abc}Y_d
$$
$$
2W_{ab;[cd]}=R^e{}_{acd}W_{eb}+R^e{}_{bcd}W_{ae}
$$
together with (\ref{a6}) and
$$
u_{a;b}={\ts{1\over3}}\Theta h_{ab}+\sigma_{ab}\,,\quad \dot{u}_a=0=
\omega_{ab}
$$
lead to the following crucial identities:
\begin{eqnarray}
\left(\d_a f\right)^{\rd} &=& \d_a \dot{f}-{\ts{1\over3}}\Theta \d_a f
-\sigma_a{}^b \d_b f
\label{a7}\\
\d_{[a} \d_{b]} f &=& 0
\label{a8}\\
\left(\d_a S_b \right)^{\rd} &=& \d_a \dot{S}_b - {\ts{1\over3}}\Theta
\d_a S_b -\sigma_a{}^c \d_c S_b +H_a{}^d\ep_{dbc}S^c
\label{a9}\\
\d_{[a} \d_{b]} S_c &=& 
\left({\ts{1\over9}}\Theta^2
-{\ts{1\over3}}\rho\right)S_{[a}h_{b]c}
-\sigma_{c[a}\sigma_{b]d}S^d 
\nonumber\\
& &{}+S_{[b}\left\{E_{c]a}-{\ts{1\over3}}\Theta
\sigma_{c]a}\right\}
+h_{c[a}\left\{E_{b]d}-{\ts{1\over3}}\Theta
\sigma_{b]d}\right\}S^d
\label{a10}\\
\left(\d_a S_{bc}\right)^{\rd} &=& \d_a \dot{S}_{bc}-{\ts{1\over3}}
\Theta \d_a S_{bc}-\sigma_a{}^d \d_d S_{bc}+2H_a{}^d\ep_{de(b}
S_{c)}{}^e
\label{a11}\\
\left(\d^b S_{ab}\right)^{\rd}&=&\d^b\dot{S}_{ab}-{\ts{1\over3}}\Theta
\d^b S_{ab}-\sigma^{bc}\d_c S_{ab}+\ep_{abc}H^b{}_dS^{cd}
\label{a11.}\\
\d_{[a}\d_{b]}S^{cd} &=& 
2\left({\ts{1\over9}}\Theta^2
-{\ts{1\over3}}\rho\right)S_{[a}{}^{(c}h_{b]}{}^{d)}
-2\sigma_{[a}{}^{(c}\sigma_{b]e}S^{d)e}
\nonumber\\
& & {} -2S_{[a}{}^{(c}\left\{E_{b]}{}^{d)}-{\ts{1\over3}}\Theta
\sigma_{b]}{}^{d)}\right\} \nonumber \\
& &{}+2h_{[a}{}^{(c}\left\{E_{b]e}-{\ts{1\over3}}\Theta
\sigma_{b]e}\right\}S^{d)e}
\label{a12}
\end{eqnarray}
Then (\ref{a7})--(\ref{a12}) imply the further important identities
\begin{eqnarray}
\ep_{abc}S^b{}_p\,\c V^{cp} &=& 2S^{bc}\d_{[a}V_{b]c}-
{\ts{1\over2}}S_{ab}\d_c V^{bc}
\label{a13} \\
\c\left(fS_{ab}\right)&=&f\c S_{ab}+\ep_{cd(a}S_{b)}{}^d\d^c f
\label{a13.}\\
\d^b\,\c S_{ab} &=& {\ts{1\over2}}\ep_{abc}
\d^b\left(\d_d S^{cd}\right)+\ep_{abc}S^b{}_d\left(
{\ts{1\over3}}\Theta\sigma^{cd}-E^{cd}\right) \nonumber \\
& &{}-\sigma_{ab}\ep^{bcd}\sigma_{ce}S^e{}_d
\label{a14} \\
\left(\c S_{ab}\right)^{\rd} &=& \c \dot{S}_{ab}-{\ts{1\over3}}\Theta
\c S_{ab} \nonumber \\
& &{}-\sigma_e{}^c\ep_{cd(a}\d^e S_{b)}{}^d+3H_{c<a}S_{b>}{}^c
\label{a15}\\
\c\c S_{ab}&=&-\d^2S_{ab}+{\ts{3\over2}}\d_{<a}\d^cS_{b>c}
+\left(\rho-{\ts{1\over3}}\Theta^2\right)S_{ab} \nonumber\\
&&{}+3S_{c<a}\left\{E_{b>}{}^c-{\ts{1\over3}}\Theta\sigma_{b>}{}^c
\right\}+\sigma_{cd}S^{cd}\sigma_{ab}\nonumber\\
&&{}-S^{cd}\sigma_{ca}\sigma_{bd}
+\sigma^{cd}\sigma_{c(a}S_{b)d}
\label{a16}\\
\d^2\left(\d_a f\right) &=& \d_a\left(\d^2 f\right)+R^*_a{}^b\d_b f
\label{a17}
\end{eqnarray}
where $\d^2=\d^a \d_a$ is the covariant Laplacian.

\end{document}